\documentclass[prl, twocolumn, superscriptaddress]{revtex4-1}
\usepackage{bm, amsmath, amsfonts, amssymb}
\usepackage[italicdiff]{physics}
\usepackage{braket}
\usepackage{comment}
\usepackage{subfigure}
\usepackage{color}
\usepackage{graphicx}

\usepackage{qcircuit}
\usepackage{url}
\usepackage{bm}
\usepackage[breaklinks=true]{hyperref}

\begin{document}
\title{Bosonic Andreev bound state}

\author{Nobuyuki Okuma}
\email{okuma@hosi.phys.s.u-tokyo.ac.jp}
\affiliation{
 Graduate School of Engineering, Kyushu Institute of Technology, Kitakyushu 804-8550, Japan
}

\date{\today}
\begin{abstract}
A general free bosonic system with a pairing term is described by a bosonic Bogoliubov-de Gennes (BdG) Hamiltonian. The representation is given by a pseudo-Hermitian matrix, which is crucially different from the Hermitian representation of a fermionic BdG Hamiltonian. 
In fermionic BdG systems, a topological invariant of the whole particle (hole) bands can be nontrivial, which characterizes the Andreev bound states (ABS) including Majorana fermions. 
In bosonic cases, on the other hand, the corresponding topological invariant is thought to be trivial owing to the stability condition of the bosonic ground state.
In this Letter, we consider a two-dimensional model that realizes a bosonic analogy of the ABS. The boundary states of this model are located outside the bulk bands and are characterized by a nontrivial Berry phase (or polarization) of the hole band. Furthermore, we investigate the zero-energy flat-band limit in which the Bloch Hamiltonian is defective, where the particle and hole states are identical to each other. In this limit, the Berry phase is $\mathbb{Z}_2$ quantized thanks to an emergent parity-time symmetry. This is an example of a topological invariant that uses the defective nature as a projection structure. 
Thus, boundary states in our model are essentially different from Hermitian topological modes and their variants.

\end{abstract}

\maketitle
The bosonic excitations from a Bose-Einstein condensate are well described by a quadratic Hamiltonian called bosonic Bogoliubov-de Gennes (BdG) Hamiltonian \cite{bogoliubov1947theory,Colpa-78,kawaguchi2012spinor,Altland-Simons}.
As well as the systems that consist of bosons such as photons \cite{ozawa2019topological,braunstein2005quantum} and bosonic atoms \cite{kawaguchi2012spinor}, the bosonic BdG Hamiltonian can describe emergent bosonic quasiparticles in ordered states such as magnons \cite{topo-magnon} and phonons \cite{topologicalphonon}.
Unlike the fermionic counterpart, its excitation spectrum is related to the eigenspectrum of a pseudo-Hermitian Hamiltonian matrix with a particle-hole symmetry \cite{BdGsym} if there exists a paring term, which breaks the particle-number conservation.
This is an example of the non-Hermitian system whose non-Hermiticity originates not from an open quantum nature but from the linear approximation of a non-linear equation.

Recently, a lot of concepts in topological physics \cite{Zhang-review,Kane-review} have been generalized to bosonic BdG Hamiltonians even though the representation matrix is pseudo-Hermitian \cite{topo-magnon}.
For example, the Chern number is defined by using a para-unitary matrix, and it characterizes the bulk-boundary correspondence \cite{Shindou-13} as in the case of Hermitian topological physics \cite{hatsugai1993chern}. Similar generalizations for other topological numbers such as $\mathbb{Z}_2$ invariant have been extensively studied \cite{Kondo-Akagi-Katsura-19}.
In the language of non-Hermitian topological physics \cite{ashida2020non, okuma2023non,lin2023topological}, this is a manifestation of the line-gap topology, which is adiabatically connected to the Hermitian topology without closing the gap and changing the symmetry \cite{KSUS-19}.

One interesting direction is to seek topological boundary states that reflect the BdG nature.
In fermionic cases, a topological number of the whole particle (hole) bands can be non-trivial, and it describes the Andreev bound states (ABS) including Majorana fermions \cite{tanaka-sato-nagaosa-11,sato-fujimoto-16}. 
In bosonic cases, on the other hand, the corresponding topological invariant can not be nontrivial if we limit our discussion to a topological phase transition in Kitaev's periodic table \cite{Schnyder-08,kitaev2009periodic}, which requires a non-trivial band inversion process. For the stability of the ground state, the bosonic excitation energies should be nonnegative. 
Owing to this stability condition, the BdG Hamiltonian is adiabatically connected to a trivial Hamiltonian without closing the gap between the particle and hole bands \cite{Shindou-13}.

In this Letter, we investigate a BdG Hamiltonian on the two-dimensional square lattice and find a bosonic analogy of ABS that is induced by a non-trivial Berry phase of the particle (hole) bands defined in an unconventional manner.
In an extreme limit, the Berry phase is $\mathbb{Z}_2$-quantized owing to an emergent parity-time symmetry.
This quantization can be understood as the non-Hermitian topology that uses the defective nature as a projection structure.

\paragraph{Basics of bosonic BdG Hamiltonian.---}
First, we review the basic properties of BdG Hamiltonians and define some notations.
A translation-invariant lattice BdG Hamiltonian with $N$ internal degrees of freedom is given by \cite{Shindou-13}
\begin{align}
    \hat{H}&=\frac{1}{2}
    \sum_{\bm{k}}(\bm{a}^{\dagger}_{\bm{k}},\bm{a}_{-\bm{k}}) H_{\bm{k}}
    \begin{pmatrix}
        \bm{a_{\bm{k}}}\\
        \bm{a}^{\dagger}_{-\bm{k}}
    \end{pmatrix},
    \label{freebdg}
\end{align}
where $\bm{a}^{\dagger}_{\bm{k}}=(a^{\dagger}_{1,\bm{k}},\cdots,a^{\dagger}_{N,\bm{k}})$ denote creation operators of bosons with crystal momentum $\bm{k}$.
$H_{\bm{k}}$ is a $2N\times 2N$ Hermitian matrix with the following form \cite{Shindou-13}:
\begin{align}
    H_{\bm{k}}=
    \begin{pmatrix}
    h_{\bm{k}}&s_{\bm{k}}\\
    s^{*}_{-\bm{k}}&h^*_{-\bm{k}}
    \end{pmatrix},
\end{align}
where $h_{\bm{k}}$ and $s_{\bm{k}}$ are $N\times N$ matrices that represent the normal term and pairing term (anomalous term), respectively.
It is well known that the excitation spectrum of the bosonic BdG Hamiltonian is not given by the eigenspectrum of $H_{\bm{k}}$ if the pairing term is nonzero, unlike in the case of fermions.
Interestingly, the true excitation spectrum is related to the eigenspectrum of a pseudo-Hermitian matrix with a particle-hole symmetry \cite{BdGsym}:
\begin{align}
    &H^{\sigma}_{\bm{k}}:=\sigma_z H_{\bm{k}},\\
    &\sigma_z[H^{\sigma}_{\bm{k}}]^{\dagger}\sigma_z=H^{\sigma}_{\bm{k}},\label{pseudo}\\
    &\sigma_x[H^{\sigma}_{-\bm{k}}]^{*}\sigma_x=-H^{\sigma}_{\bm{k}}\label{phs},
\end{align}
where $\sigma$'s denote the Pauli matrices in Nambu space.
This pseudo-Hermitian matrix is diagonalized by a paraunitary matrix $P_{\bm{k}}$ \cite{Shindou-13}:
\begin{align}
    &P_{\bm{k}}^{-1}H^{\sigma}_{\bm{k}}P_{\bm{k}}=
    \begin{pmatrix}
        E_{\bm{k}}&0\\
        0&-E_{-\bm{k}}
    \end{pmatrix},\\
    &P_{\bm{k}}\sigma_zP^{\dagger}_{\bm{k}}=P^{\dagger}_{\bm{k}} \sigma_z P_{\bm{k}}=\sigma_z.\label{paraunitary}
\end{align}
Here, $E_{\bm{k}}$ is the diagonal matrix whose elements $\{\epsilon_{\bm{k},a}~|~a=1,\cdots,N\}$ are the excitation energies.
Note that the excitation energies can be negative or complex without further assumptions.
Since the former/latter leads to the Landau/dynamical instability of the ground state \cite{kawaguchi2012spinor}, the positive semidefiniteness of the Hermitian matrix $H_{\bm{k}}$ is assumed to realize the nonnegative excitation energies in conventional condensed matter physics \footnote{In some isolated cold atomic systems, negative- or complex-energy states are relevant in physics. See Refs.\cite{kawaguchi2012spinor,Ohashi-Kobayashi-Kawaguchi-20}. }.
In the following, we call $\{\epsilon_{\bm{k},a}\}$ and $\{\epsilon_{\bm{k},-a}:=-\epsilon_{-\bm{k},a}\}$ the particle and hole bands, respectively.
Owing to the non-Hermiticity (non-normality), the bra (left) eigenvectors are not always the hermitian conjugate of the ket (right) eigenvectors:
\begin{align}
    \langle\!\langle \bm{k},i|H^{\sigma}_{\bm{k}}=\epsilon_{\bm{k},i}\langle\!\langle \bm{k},i|,\notag\\
    H^{\sigma}_{\bm{k}}\ket{\bm{k},i}=\epsilon_{\bm{k},i}\ket{\bm{k},i},
\end{align}
where $i$ takes both $a$ and $-a$.
If we take the biorthonormal convention, the bra and ket eigenvectors are in the following relations \cite{Ohashi-Kobayashi-Kawaguchi-20,KSUS-19}:
\begin{align}
    &|\bm{k},i\rangle\!\rangle=\mathrm{sgn}(i)\sigma_z|\bm{k},i\rangle,\label{biortho1}\\
    &\langle\!\langle \bm{k},i|\bm{k},j\rangle=\mathrm{sgn}(i)\bra{\bm{k},i}\sigma_z|\bm{k},j\rangle=\delta_{i,j}.\label{biortho2}
\end{align}

In topological physics of bosonic BdG systems, the topological invariants are usually defined by using both the right and left eigenvectors.
For example, the Berry connection defined in Ref. \cite{Shindou-13,Ohashi-Kobayashi-Kawaguchi-20,engelhardt2015topological} is rewritten as follows:
\begin{align}
    A^{LR}_{i,\nu}(\bm{k}):&=i \mathrm{Tr}[\Gamma_i\sigma_z P^{\dagger}_{\bm{k}}\sigma_z(\partial_{k_\nu}P_{\bm{k}})]
    =i \mathrm{Tr}[\Gamma_i P^{-1}_{\bm{k}}(\partial_{k_\nu}P_{\bm{k}})]\notag\\
    &=i\langle\!\langle \bm{k},i|\partial_{k_\nu}\ket{\bm{k},i},
\end{align}
where $\Gamma_i$ is a diagonal matrix taking $+1$ for the $i$-th diagonal component and zero otherwize.
We have used the para-unitary condition (\ref{paraunitary}) and assumed the biorthonormal convention. The Chern number is defined by using $A^{LR}_{i,\nu}$, which describes the topological physics of the bulk-boundary correspondence. 
As mentioned in the introduction part, these topological invariants cannot be nontrivial for the whole particle (hole) bands if we assume the positive definiteness of $H_{\bm{k}}$, which ensures the positivity of the excitation energies. This is because the Hamiltonian $H^{\sigma}_{\bm{k}}$ under this condition is adiabatically connected to $1_{N \times N}\otimes\sigma_z$ without closing the gap between the particle and hole bands \cite{Shindou-13}.
In the following, we seek another possibility: the boundary states induced by the polarization of the particle (hole) bands.

\begin{figure*}
\begin{center}
 \includegraphics[width=17cm,angle=0,clip]{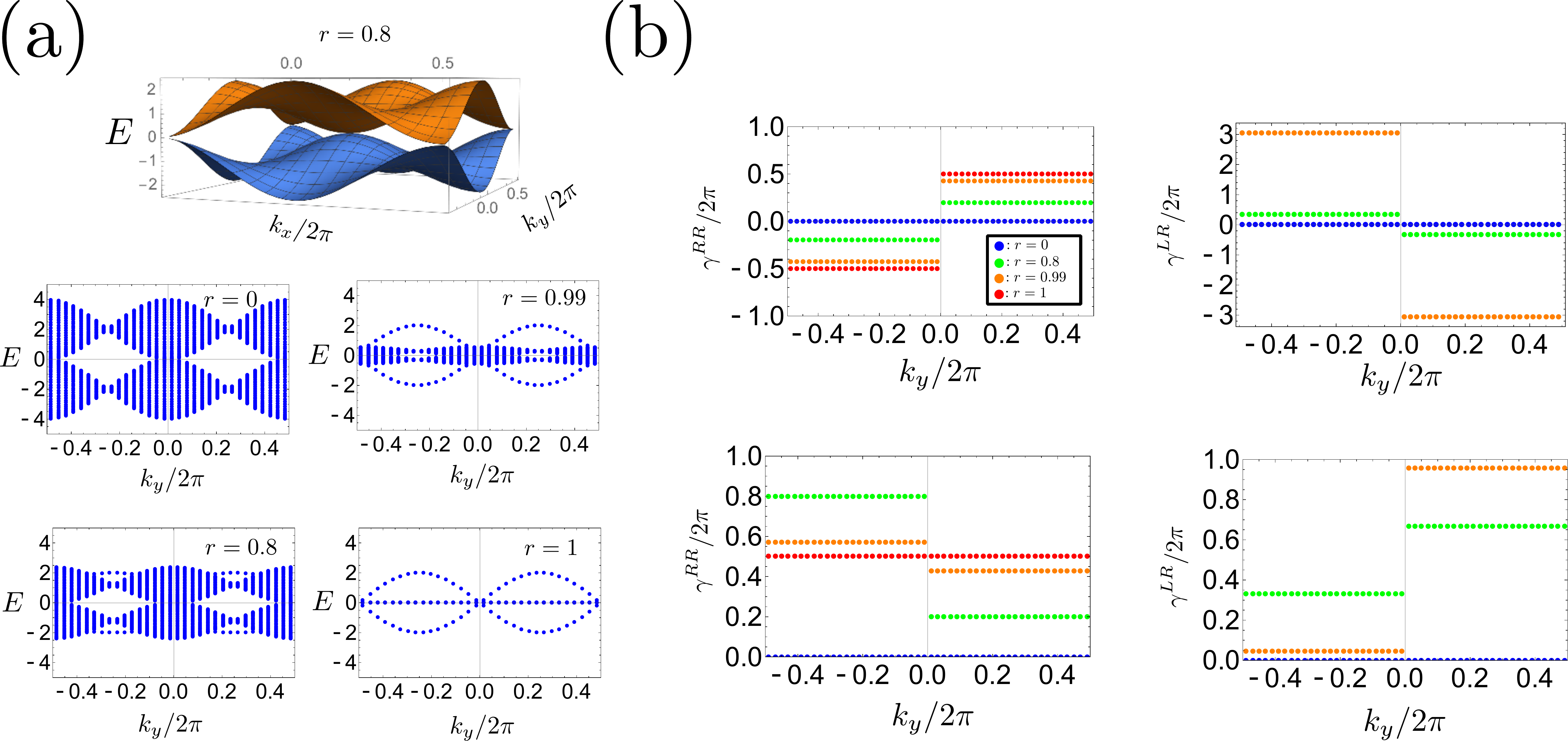}
 \caption{(a) Band structure for $r=0.8$ and $k_y$-resolved dispersion on a cylinder for various $r$. The system size is $32\times 32$. A $\pi$ flux is inserted in the cylinder to avoid $k_y=0,\pi$. (b) Berry phases $\gamma^{RR}$ and $\gamma^{LR}$ for various $r$. Upper panel: calculation using a Bloch wave that is a continuous function of $\bm{k}$. Lower panel: the remainder of the Berry phase divided by $2\pi$.  The momenta $k_y=0,\pi$ are avoided. The system size is $200\times50$. For numerical integration, the size in the $x$ direction is taken much larger than that in the $y$ direction.   }
 \label{fig1}
\end{center}
\end{figure*}

\paragraph{Model with nontrivial Berry phase.---}
According to the ``modern theory" of polarization \cite{spaldin2012beginner}, the bulk polarization is given by the Berry phase (divided by $2\pi$) that is defined as the integration of the Berry connection on a non-contractible loop in the Brillouin zone.
In one dimension, the Berry phase is given by
\begin{align}
    &\gamma_i=i\int^{\pi}_{-\pi} dk~ \bra{k,i}\partial_k\ket{k,i},
\end{align}
where $k$ is the one-dimensional crystal momentum with lattice constant $a=1$. 
In recent topological physics, bulk polarization is known as another route to induce the boundary states.
We here generalize this idea to the particle (hole) bands of a bosonic BdG Hamiltonian.

Let us consider the following two-dimensional bosonic BdG Hamiltonian:
\begin{align}
    H^{\sigma}_{\bm{k}}=&2(1-\cos k_x\cos k_y)\sigma_z\notag\\
    &+2ir\left[(\cos k_y-\cos k_x)\sigma_x+\sin k_x \sin k_y\sigma_y\right],\label{bloch}
\end{align}
where $r\geq0$ describes the strength of the paring term.
This Hamiltonian satisfies Eqs. (\ref{pseudo}) and (\ref{phs}).
The eigenspectrum is given by
\begin{align}
    E_{\pm}(\bm{k})=\pm2\sqrt{1-r^2}(1-\cos k_x\cos k_y),
\end{align}
where $\pm$ denotes particle and hole bands [Fig. \ref{fig1}(a)].
From this expression, we further assume $r\leq1$ to ensure the nonnegativity of particle energies, $E_{+}(\bm{k})\geq0$, which is the condition for the stable ground state. 
At the extreme limit $r=1$, the energy spectrum becomes flat.

The Bloch Hamiltonian (\ref{bloch}) describes a system with periodic boundary conditions in both the $x$ and $y$ directions.
To discuss the corresponding boundary states, we impose the open/periodic boundary condition in the $x/y$ direction.
In Fig. \ref{fig1}(a), we plot the eigenspectrum with respect to the momentum in $y$ direction, $k_y$.
While there are no isolated modes for $r=0$,
we find the isolated modes outside the bulk particle and hole bands for a large $r$. 
This behavior is very different from that of Hermitian boundary states, which are located inside the band gap.
At some $k_y$, the isolated modes are absorbed into the bulk bands.
The isolated modes do not degenerate at each momentum $k_y$ and are localized at one boundary.
The side of this localization depends on the sign of $k_y$ and the particle-hole band index.
At the extreme limit $r=1$, the low-energy dispersion of the boundary states becomes linear and gapless, and the boundary states are connected to the bulk states at momenta $k_y=0,\pi$.
In other words, the boundary states look like chiral boundary modes around these symmetric points. 

These boundary modes are induced by the bulk polarization defined for the Bloch states.
In the non-Hermitian (pseudo-Hermitian) cases, however, one can define two types of Berry connections at each momentum $k_y$:
\begin{align}
    &\gamma_i^{LR}(k_y)=i\int^{\pi}_{-\pi} dk_x \langle\!\langle \bm{k},i|\partial_{k_x}\ket{\bm{k},i},\label{lrberry}\\
    &\gamma_i^{RR}(k_y)=i\int^{\pi}_{-\pi} dk_x \langle \bm{k},i|\partial_{k_x}\ket{\bm{k},i}\label{rrberry}.
\end{align}
The former definition \cite{Ohashi-Kobayashi-Kawaguchi-20,engelhardt2015topological}, which uses both the right and left eigenvectors, reflects the conventional manner in the line-gap topology. In this definition, we have assumed the biorthonormal conventions (\ref{biortho1}) and (\ref{biortho2}).
The latter definition, which uses only the right eigenvectors, looks like the Hermitian Berry phase.
The crucial difference from the Hermitian one is that
the set of the right eigenvectors can not span the whole Hilbert space if the paring term is nonzero. In this definition, we have assumed the normalization $\bra{\bm{k},i}\bm{k},i\rangle=1$.
The Berry connections in both definitions take real values under the present biorthonormal/normal conventions. 
Owing to the gauge degree of freedom, the Berry phases (\ref{lrberry}) and (\ref{rrberry}) are determined modulo $2\pi$.
In Fig. \ref{fig1}(b), we plot two types of the Berry phases of the hole band at each $k_y$.
In the calculation, we have used a Bloch wave that is a continuous function of momentum. As $r$ is increased,  $\gamma^{RR}$ converges to $\pm \pi$ for positive/negative $k_y$. In Hermitian topological physics, the value $\pi$ indicates the emergence of the boundary states, which indicates that $\gamma^{RR}$ can characterize the observed boundary states.
Another definition $\gamma^{LR}$, on the other hand, diverges as for the increase of $r$.
Moreover, one cannot define it at the extreme limit $r=1$ because $\bra{\bm{k},i}\sigma_z\ket{\bm{k},i}=0$, which leads to the failure of the biorthonormal convention.
Thus, we conclude that $\gamma^{RR}$ is the true definition to characterize the present boundary states. 

Note that these boundary states are essentially different from the bosonic topological boundary modes in previous studies that are defined for the gap between the particle bands. In our case, the boundary states are characterized by a quantity that is defined for the gap between the particle and hole bands.
We call the present boundary states the bosonic ABS.

\paragraph{Non-Hermitian topology at flat-band limit.---}
For general $r$, the bosonic ABS is geometrical rather than topological because the Berry phase $\gamma^{RR}$ varies continuously and is adiabatically connected to zero. At the extreme limit $r=0$, however, one can find a non-Hermitian topology in the following sense. In this limit, the band becomes completely flat, and its energy is exactly zero.
Owing to the particle-hole symmetry, $\sigma_x\ket{-\bm{k},i}^*$ is an eigenstate of $H^{\sigma}_{\bm{k}}$ with an eigenenergy $-\epsilon_{-\bm{k},i}$.
In the present case, both $\ket{\bm{k},i}$ and $\sigma_x\ket{-\bm{k},i}^*$ are the zero-energy states.
Moreover, the Hamiltonian $H^{\sigma}_{\bm{k}}$ is defective (i.e. not diagonalizable), and the hole eigenstate is identical to the particle one, except for at $\bm{k}=(0,0)$ and $(\pi,\pi)$.
Thus, $\ket{\bm{k},i}$ is identical to $\sigma_x\ket{-\bm{k},i}^*$ up to the phase, which means that the particle-hole symmetry effectively acts as if the time-reversal symmetry at $r=1$. In addition, $H^{\sigma}_{\bm{k}}$ is invariant under the inversion $\bm{k}\rightarrow-\bm{k}$. In total, we can define an effective parity-time symmetry, and $\ket{\bm{k},i}$ is identical to $\sigma_x\ket{\bm{k},i}^*$ up to the phase. 
It is known that the Berry phase under this type of antiunitary symmetry is $\mathbb{Z}_2$ quantized \cite{hatsugai2006quantized}.
In the present case, it is checked by the following calculation:
\begin{align}
    \gamma_i^{RR}(k_y)&\equiv i\int^{\pi}_{-\pi} dk_x \langle \bm{k},i|^*\sigma_x\partial_{k_x}\sigma_x\ket{\bm{k},i}^* ~(\mathrm{mod}~2\pi)\notag\\
    &=-\gamma_i^{RR}(k_y).
\end{align}
From this relation, the Berry phase is quantized to $0$ or $\pi$. The bosonic ABS corresponds to $\gamma^{RR}_i=\pi$. 

Thanks to the two-band nature, the above physics is also explained by a $\mathbb{Z}$ topology of the Hamiltonian itself.
The Bloch Hamiltonian in the defective region takes the following form:
\begin{align}
    H^{\sigma}_{\bm{k}}\propto 
    \begin{pmatrix}
        1&ih^*_{\bm{k}}\\
        ih_{\bm{k}}&-1
    \end{pmatrix},
\end{align}
where $h_{\bm{k}}\in\mathbb{C}$ is on the unit circle in the complex plane.
In other words, the degree of freedom of the Bloch Hamiltonian is limited to U(1) if we impose that the matrix is defective.
Thus, one can define the winding number on a closed loop $C$ in the Brillouin zone:
\begin{align}
    W=\frac{1}{2\pi i}\oint_C d \ln h.\label{wnumber}
\end{align}
This winding number (\ref{wnumber}) characterizes two related properties.
First, it describes the polarization at each $k_y$:
\begin{align}
    W(k_y)=\frac{1}{2\pi i}\int_{-\pi}^{\pi}dk_x \frac{d\ln h}{dk_x}=\int_{-\pi}^{\pi}\frac{dk_x}{2\pi}\theta_{\bm{k}},
\end{align}
where $\theta_{\bm{k}}=\arg (h_{\bm{k}})$.
At special points $k_y=\pm \pi/2$, the winding number is $\pm1$, which is easily calculated by
$\theta_{\bm{k}}=\pm k_x$. At general $k_y$ except for $k_y=0,\pi$, $W(k_y)=\mathrm{sgn}(k_y)$.
Remarkably, $\mathbb{Z}$ topological number can distinguish $\pm1$, while the Berry phase cannot.
Second, the winding number (\ref{wnumber}) detects the non-defective points in the Brillouin zone.
Near the non-defective point $\bm{k}=(0,0)$, the Hamiltonian takes the following form:
\begin{align}
    H^{\sigma}_{\bm{k}}&\simeq (k_x^2+k_y^2)\sigma_z+i(k_x^2-k_y^2)\sigma_x+2ik_xk_y\sigma_y\notag\\
    &\propto 
    \begin{pmatrix}
        1&ie^{-i2\phi}\\
        ie^{i2\phi}&-1
    \end{pmatrix},
\end{align}
where $\bm{k}=k(\cos\phi,\sin\phi)$ with $k=|\bm{k}|$. Thus, $\bm{k}=(0,0)$ is characterized by $W=+2$. Similarly, $\bm{k}=(\pi,\pi)$ is characterized by $W=-2$. The total winding number around non-defective points is 0, which is a non-Hermitian analogy of the Nielsen-Ninomiya theorem \cite{nielsen1981no}. 
Note that the paring term is proportional to an effective model of the quadratic band touching in Hermitian topological physics. The connection between our non-Hermitian topology and the physics of quadratic band touching, including the geometry-induced surface states \cite{yang2014emergent} and the Euler number \cite{ahn2019failure}, may be an interesting future work.
Note also that the winding number itself can be defined for general $r$ except for $0$. In that case, however, the degree of freedom of the Hamiltonian is larger than $U(1)$.

\paragraph{Discussion.---}
We here investigate several remaining topics and discuss related future works.
The bosonic ABS is expected to be found in various dimensions. In one dimension, however, the realization of a model with one positive- and one negative-energy boundary states localized at opposite boundaries seems to be difficult. If possible, these boundary states are related to each other by the particle-hole symmetry, which does not act on real-space coordinates. Thus, the boundary states are localized at the same boundary, which conflicts with the localization at the opposite boundaries.
In our two-dimensional model, the particle-hole symmetry is absent at each $k_y$ except for $0,\pi$, and the localization at the opposite boundary is allowed in each momentum sector.
Instead, one can consider the bosonic ABS in a model with a unitary symmetry. 
For example, let us consider the following one-dimensional model:
\begin{align}
    H^{\sigma}_k=1_{2\times2}\otimes\sigma_z+ir(\cos k\tau_x+\sin k\tau_y)\otimes\sigma_y,
\end{align}
where $\tau$'s are the Pauli matrices in orbital space, and $0\leq r\leq 1$. This model commutes with $\tau_z\otimes\sigma_z$ and is block-diagonalized into $\tau_z\otimes\sigma_z=\pm 1$ sectors. Each block is given by
\begin{align}
    \begin{pmatrix}
        1&ire^{\mp ik}\\
        ire^{\pm ik}&-1
    \end{pmatrix},
\end{align}
which means the non-trivial polarization. 
A generalization of the bosonic ABS may be an interesting future work. Since particle and hole states coincide at an extreme parameter, $\mathbb{Z}_2$ nature is expected to be essential if we impose that the total topological number of the particle and hole bands is zero.

Another interesting issue is the entanglement property of the ground state of a model with the bosonic ABS.
In free bosonic systems, the entanglement entropy of the ground state is induced only by the paring term. Such a term is naturally introduced by squeezed states of light and plays an important role in continuous-variable (CV) quantum computing \cite{braunstein2005quantum}.
In our calculation \footnote{See Supplemental Material.}, boundary states similar to those in our model (but with degeneracy) are found in the CV surface codes (CVSC) \cite{zhang2008anyon,milne2012universal,morimae2013continuous,demarie2014detecting}.
Reference \cite{demarie2014detecting} claims that topological entanglement entropy \cite{kitaev2006topological,levin2006detecting} is not quantized in a physical CVSC.
In Supplemental Material, we investigate the entanglement structure of the ground state of the model (\ref{bloch}) by using the formula in Ref. \cite{demarie2012pedagogical} and find behaviors similar to those in CVSC. A relationship between CV topological order and the bosonic ABS is also an interesting remaining topic.

\begin{acknowledgements}
I thank Tomonari Mizoguchi for the fruitful discussions.
This work was supported by JSPS KAKENHI Grant No.~JP20K14373 and No.~JP23K03243.
\end{acknowledgements}

\bibliography{NH-topo}

\widetext
\pagebreak

\renewcommand{\theequation}{S\arabic{equation}}
\renewcommand{\thefigure}{S\arabic{figure}}
\renewcommand{\thetable}{S\arabic{table}}
\setcounter{equation}{0}
\setcounter{figure}{0}
\setcounter{table}{0}

\begin{center}
{\bf \large Supplemental Material for 
``Bosonic Andreev bound state"}
\end{center}

\begin{figure}[b]
\begin{center}
 \includegraphics[width=16cm,angle=0,clip]{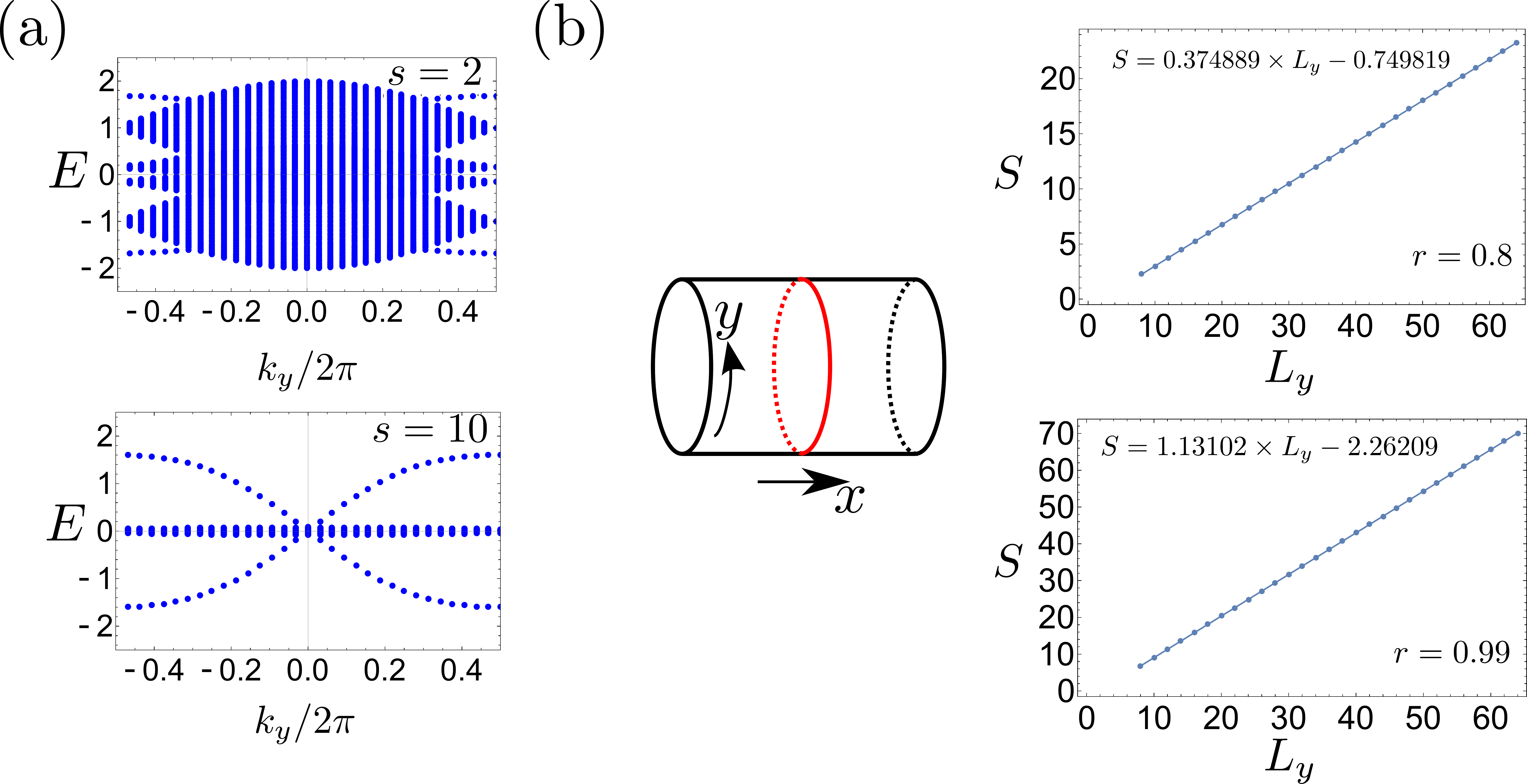}
 \caption{(a) The Bogoliubov spectrum of the physical CVSC \cite{demarie2014detecting} on the cylinder. $s$ is the squeezing parameter. The system size is $32\times32$. (b) Size dependence of the entanglement entropy of the ground state of the model (\ref{ourmodel}) on the cylinder.}
 \label{supfig}
\end{center}
\end{figure}

\section{Entanglement entropy of a model with bosonic Andreev bound states}
The physical continuous-variable surface code (CVSC) \cite{demarie2014detecting}, which is a generalization of the CVSC \cite{zhang2008anyon,milne2012universal,morimae2013continuous}, is characterized by the ``topological" entanglement entropy whose value is a continuous function of a system parameter \cite{demarie2014detecting}.
In our calculation [Fig.\ref{supfig} (a)], this model (with $45^{\circ}$ rotation from the original one, $4$ sites in 1 unit cell) has boundary modes similar (but with degeneracy) to our model analyzed in the main text.
Motivated by this fact, we investigate the entanglement entropy of our model.

Our model on torus configuration is restated as follows:
\begin{align}
    &\hat{H}=\frac{1}{2}
    \sum_{\bm{k}}(\bm{a}^{\dagger}_{\bm{k}},\bm{a}_{-\bm{k}}) H_{\bm{k}}
    \begin{pmatrix}
        \bm{a_{\bm{k}}}\\
        \bm{a}^{\dagger}_{-\bm{k}}
    \end{pmatrix}=\frac{1}{2}
    \sum_{\bm{k}}(\bm{a}^{\dagger}_{\bm{k}},-\bm{a}_{-\bm{k}}) H^{\sigma}_{\bm{k}}
    \begin{pmatrix}
        \bm{a_{\bm{k}}}\\
        \bm{a}^{\dagger}_{-\bm{k}}
    \end{pmatrix},\notag\\
   &H^{\sigma}_{\bm{k}}=2(1-\cos k_x\cos k_y)\sigma_z
    +2ir\left[(\cos k_y-\cos k_x)\sigma_x+\sin k_x \sin k_y\sigma_y\right],\label{ourmodel}
\end{align}
where $0\leq r\leq 1$.
We here investigate the cylindrical configuration ($x$: open,~$y$: periodic) and measure the entanglement entropy $S$ of the ground state for half of the system with $L_x=2L_y$. 
A formula for the entanglement entropy of the quantum harmonic oscillator, which is identical to the bosonic BdG system, is given in Ref. \cite{demarie2012pedagogical}.
Using this formula with the Fourier transform, we calculate the size dependence of the entanglement entropy [Fig. \ref{supfig}(b)]. While the dominant term obeys the area law [i.e., $S_{\rm dom}\propto L_y$], there is a negative constant subleading term. The value depends on the model parameter $r$, which is similar to the behavior of the physical CVSC.
In gapped systems, this type of subleading term is called topological entanglement entropy because it characterizes the topological order \cite{kitaev2006topological,levin2006detecting}.
Unlike in the gapped systems, the constant subleading term is not quantized in the physical CVSC and our model.

\end{document}